\documentclass[twocolumn,aps,prc,superscriptaddress]{revtex4-2}

\usepackage{color,amsmath,amssymb,epsfig,graphicx}
\usepackage{bm}
\usepackage{xcolor}
\usepackage{mathptmx, courier, pifont}
\usepackage[scaled=0.92]{helvet}
\usepackage[T1]{fontenc}
\usepackage{textcomp}
\usepackage[colorlinks=true,linkcolor=blue,filecolor=blue,urlcolor=blue,citecolor=blue]{hyperref}
\providecommand{\boldsymbol}[1]{\mbox{\boldmath $#1$}}

\usepackage{multirow}
\usepackage{longtable}
\usepackage[utf8]{inputenc}
\usepackage{array}
\usepackage{booktabs}
\usepackage{textgreek}
\usepackage{xcolor}
\renewcommand{\arraystretch}{1.6}
\begin{document}
\author{Saliha Tak}
\affiliation{Department of Physics, Islamic University of Science and Technology, Kashmir -192122, INDIA}
\author{N. Rather}
\email{qadriniyaz@gmail.com}
\affiliation{Department of Physics, Islamic University of Science and Technology, Kashmir -192122, INDIA}
\author{J. A. Sheikh}
\email{sjaphysics@gmail.com}
\affiliation{Department of Physics, University of Kashmir, Srinagar-190004 INDIA}
\author{S. P. Rouoof}
\affiliation{Department of Physics, Islamic University of Science and Technology, Kashmir -192122, INDIA}
\author{S. Jehangir}
\affiliation{Department of Physics, Govt. Degree College Kulgam, Jammu and Kashmir, 192231, INDIA}
\author{G. H. Bhat}
\affiliation{Department of Physics, GDC Shopian, Higher Education, Jammu and Kashmir, 192303,, INDIA}

\title{ Isospin-invariant Skyrme density functional investigation of quadrupole deformations in the isospin triplets
  of $\bf{A=70, 74}$ and $\bf{78}$ }

\begin{abstract}
  
  In this work, we perform self-consistent isospin-invariant Skyrme density functional calculations to investigate the ground-state properties of the isospin
  triplets with mass numbers $A=70$, $74$, and $78$. The present study is motivated by  
  the reported anomaly between the observed quadrupole deformation and that expected from isospin symmetry. It is shown that the trend of the 
  deduced $T_z$ dependence of $B(E2,2_1^+ \rightarrow 0_1^+)$ transition strengths from the charge
  quadrupole moments is consistent with the measured values for the $T_z=\pm 1$ 
  members of the $A=70$ triplet. However, the calculated $B(E2,2_1^+ \rightarrow 0_1^+)$ rates for the $T_z=0$ members exceed
  the experimental transition rates in all the three studied triplets. Our study implies that non-linear dependence of the $B(E2,2_1^+ \rightarrow 0_1^+)$ transition strengths is expected in all of them and  
  primarily arises from the observed suppressed transition rates for the $T_z=0$ members. 

\end{abstract}

\date{\today}

\maketitle
\section{Introduction}
\label{Sect.01}

The isospin symmetry, introduced by Heisenberg in 1932 \cite{Heisenberg1932}, has played an invaluable role in elucidating the
properties of atomic nuclei \cite{Bentley2007,Sheikh2024Isospin}. 
It was proposed based on the similar properties of neutrons and protons in the absence of electromagnetic interaction, and was formulated along
the lines of the SU(2) algebra of the spin angular momentum. The rotations are defined in the abstract isospin space and considering that proton and neutron
parts of the total nuclear Hamiltonian are same, it remains invariant under rotations in this space. This leads to the
conservation of the isospin quantum number, $T$.
This symmetry gives rise to isobaric analog states (IAS) \cite{Bentley2007}, corresponding to different $T_z$ for a given
$T$. In the limit of exact isospin symmetry, the IAS are supposed to exhibit identical nuclear properties that include excitation spectra
and electromagnetic transition probabilities. The most studied among these properties is the excitation spectra of the mirror nuclei, which
depict a remarkable similarity after correcting for the Coulomb energy differences \cite{BohrMottelson1969}.

Nevertheless, the isospin symmetry is broken in atomic nuclei and its primary cause is the Coulomb interaction among protons. The strong interaction
also breaks the isospin symmetry and is directly evident from the differences in the  neutron-neutron, neutron-proton
and proton-proton $^{1}S_0$ scattering lengths. This breaking mainly arises from the up- and down-quark mass splitting 
and the electromagnetic interaction among quarks and gluons \cite{Miller2006}. The isospin symmetry breaking effects
manifest themselves through Coulomb displacement
energies \cite {deAngelis2001Coulomb}, deviations from the
quadratic form of the isobaric multiplet mass equation \cite{Sheikh2024Isospin,Bentley2007,Dong2018,ORMAND19891} and Nolen-Schiffer anomaly
\cite{Nolen1969,Auerbach1983}. Although the excitation energies provide an important probe on the isospin symmetry breaking effects, they do
not directly test the isospin purity of the nuclear wave function.

Electromagnetic transition matrix elements provide a more stringent probe of the isospin symmetry. In particular, electric quadrupole ($E2$) transition strengths probe the
spatial distribution of protons and are, therefore, sensitive to both intrinsic deformation and isospin-dependent structural effects.
For a $T=1$ isospin triplet, charge independence of the nuclear interaction implies a simple linear dependence of the proton matrix elements
on $T_z$ \cite {Bernstein1979}.
Consequently, the $B(E2,2_1^+ \rightarrow 0_1^+)$ values for analogous transitions within an isospin triplet are expected to vary
linearly with $T_z$. Deviations from this linear behaviour would indicate changes in intrinsic structure  or breakdown of the isospin
symmetry \cite{Lenzi2009Isospin,Satula2016Isospin,Bentley2007}.

The interplay between isospin symmetry and collective motion is particularly prominent in nuclei with $N \approx Z$, where protons and neutrons occupy the same shell-model orbitals and strong proton-neutron 
correlations are expected. In the mass region $A \approx 70$-80, both experimental and theoretical
studies indicate the onset of significant quadrupole
deformation and the emergence of rotational structures \cite{Delaroche2010,Moller2016,Nichols2014,Gade2005,Iwasaki2014,Morse2018,Lemasson2012,Davies2007,NaraSingh2007,Lister1987}. For even-even nuclei, 
quadrupole collectivity is measured from the strength of the $B(E2,2^+_1 \rightarrow 0^+_1)$ transition probability, and is also inferred from
the excitation energy of the $2^+_1$ state.
These quantities are empirically related through Grodzins' rule \cite{GRODZINS196288}.
This empirical relationship implies that comparable $2^+_1$ excitation energies within an isospin triplet are expected to correspond
to similar intrinsic deformations and $B(E2,2_1^+ \rightarrow 0_1^+)$ strengths.

A recent study has reported an apparent anomaly between $B(E2,2_1^+ \rightarrow 0_1^+)$ rates and the expected linear
trend from the isospin symmetry requirement for the $A = 70$, $74$ and $78$ isospin triplets \cite{tension_Henderson}. Similar observed excitation energies of the $2^+_1$ states, which 
indicate comparable collectivity among the triplet members, appear to contradict the anticipated linear dependence of the $B(E2,2_1^+ \rightarrow 0_1^+)$ transition 
strengths on $T_z$. Further, pronounced non-linear $T_z$ dependence of $B(E2,2_1^+ \rightarrow 0_1^+)$ rates has previously 
been observed in the isobaric analog states of the isospin triplet corresponding to $A = 70$ and interpreted as a consequence of the shape differences between
the mirror partners of the isospin triplet \cite{shape_Wimmer}. These observations raise an important question regarding the robustness of isospin symmetry in deformed systems 
and suggest that the $N \approx Z \approx 40$ region may represent a unique mass domain where systematic deviations from linear $B(E2,2_1^+ \rightarrow 0_1^+)$ trends within an isospin multiplet 
might become significant. Further, a more recent analysis \cite{Backes2025}, based on the density functional
approach combined with the generalized Bohr Hamiltonian and angular momentum projection, demonstrated that shape changes between the mirror partners of the studied isospin triplet states are unlikely.

In the aforementioned studies, the isospin projection has not been performed and it
is well known that projection of isospin quantum number is essential to properly describe the nuclei having
$N \sim Z$, in particular, for odd-odd nuclei for which $T=0$ and $1$
states are mixed at the mean-field level. The purpose of the present work is to investigate the isospin triplets of  mass numbers $A=70$, $74$, and $78$
using an approximate isospin projection formulation. In this approach, referred to as the isocranking, the members of an
isobaric chain are obtained by rotation in the isospin space \cite{Sato_2013,Sheikh_2014}. 

In the present work, we shall employ the nuclear density functional approach which is one of the most powerful methods to describe atomic nuclei all across the
\textnormal{Segr\`{e}} chart \cite{Bender2003} and it is
well established to describe the ground-state properties with remarkable accuracy. For instance, it is now possible to reproduce the binding energies
with root-mean-square deviation of less than $0.6$ MeV \cite{Goriely2009}. Nevertheless, it is known that the conventional density functional theory (DFT) approaches
in atomic nuclei
are inadequate to describe the properties of nuclei close to $N=Z$. For instance, the Wigner energy term is incorporated in an $ad$ $hoc$ manner to
reproduce the binding energies of nuclei in the vicinity of $N=Z$ \cite{Samyn2003}. The main reason for this discrepancy arises due to the neglect of the neutron-proton
mixed densities in the conventional formulation of nuclear DFT. 
The density functional includes only isoscalar and $T_z$ component of the isovector densities, the $T_x$ and $T_y$ components
are disregarded \cite{Bender2003}. This simplification is questionable
in lighter and medium-mass nuclei, where protons and neutrons occupy the same shell-model orbitals and 
proton–neutron correlations play a central role. 

To address the above limitation, generalized isospin-invariant energy density functional framework has been developed that explicitly incorporates 
proton–neutron mixed densities in the functional \cite{Perlinska2004,Sato_2013,Sheikh_2014}. In this approach, the nuclear energy density functional is constructed from the isoscalar and the three components of the
isovector densities. The new development provides a natural theoretical framework for investigating isobaric analog states in nuclei
near the $N = Z$ line. Using this generalized DFT framework, we have performed fully self-consistent calculations for the $A=70, 74$ and $78$
isospin triplets. For the $T_z=\pm 1$ systems, the normal DFT calculations have been performed by constraining proton and neutron particle numbers.
In the case of $T_z=0$ systems, the isocranking has been used to obtain the self-consistent mean-field solutions. The manuscript is organized in the
following manner. In the next section, the cranking Skyrme DFT approach is briefly outlined and in section III, the results are discussed for  the
three chains of $A=70, 74$ and $78$. Finally, the summary and conclusions are presented in section IV.

\section{Isospin-Invariant Skyrme Density Functional Approach}

In the present work, nuclear DFT is employed in an isospin-invariant formulation that allows for proton-neutron mixing at the
mean-field level. In this framework, the Kohn--Sham reference state is taken to be a single Slater determinant constructed from fully occupied single-particle
states,
\begin{equation}
|\Phi\rangle = \prod_{k=1}^{A} c_k^\dagger |0\rangle ,
\end{equation}
where the creation operators $c_k^\dagger$ correspond to fermionic single-particle operators depending on spatial, spin, and isospin coordinates, and are related
to the Kohn--Sham orbitals through a unitary transformation \cite{Sato_2013,Sheikh_2014}. In contrast to the conventional implementations of nuclear DFT, the single-particle states are allowed
to contain both neutron and proton components, thereby permitting explicit neutron--proton mixing in the particle--hole channel. As a result of neutron-proton mixing, the DFT 
formalism implements the isospin projection technique to project the states of good $T_z$ corresponding to the physical nuclei of an isospin multiplet. As discussed in the previous
section, this generalization is essential for a correct microscopic description of nuclei close to the $N=Z$ line, where neutrons and protons occupy the same
shell-model orbitals and isovector correlations play a significant role.

The local densities entering the energy density functional are obtained from the generalized one-body density matrix and can be decomposed into isoscalar and
isovector components. The total energy of the nucleus is expressed as
\begin{equation}
E[\rho] = \int d^3r \, \mathcal{H}(\boldsymbol{r}),
\end{equation}
where the Hamiltonian density $\mathcal{H}(\boldsymbol{r})$ is constructed from bilinear combinations of local particle, kinetic, spin, isospin, current, spin--current
and isospin-current densities, in both the isoscalar and isovector channels \cite{Perlinska2004}. In the limit of exact isospin symmetry, the functional is invariant under global rotations in isospin
space.

To incorporate explicit isospin symmetry breaking (ISB), the Hamiltonian density is written as the sum of isospin-invariant and isospin-breaking contributions,
\begin{equation}
\mathcal{H}(\boldsymbol{r})
=
\mathcal{H}_{\mathrm{Sk}}(\boldsymbol{r})
+
\mathcal{H}_{\mathrm{Coul}}(\boldsymbol{r})
+
\mathcal{H}_{\mathrm{ISB}}(\boldsymbol{r}),
\end{equation}
where $\mathcal{H}_{\mathrm{Sk}}$ denotes the isospin-conserving Skyrme energy density functional, $\mathcal{H}_{\mathrm{Coul}}$ represents the Coulomb interaction, and
$\mathcal{H}_{\mathrm{ISB}}$ accounts for nuclear ISB effects beyond the Coulomb force. The latter includes both charge-symmetry-breaking (CSB) and
charge-independence-breaking (CIB) contributions, which arise from the underlying strong interaction. The inclusion of these terms enables a unified and consistent
microscopic treatment of electromagnetic and nuclear sources of isospin symmetry breaking.

Since the self-consistent solutions of the isospin-invariant DFT equations generally break isospin symmetry spontaneously at the mean-field level, states with good $T_z$ need to be extracted. In order to obtain these states
corresponding to the different members of an isobaric multiplet, an approximate restoration of isospin symmetry is achieved using an isocranking-based isospin
projection method, following Refs.\cite{Sato_2013,Sheikh_2014}. In this approach, the following auxiliary Kohn--Sham equation is solved:

\begin{align}
  \widehat{h}^\prime \phi_k=\epsilon_k \phi_k;
\end{align}

\begin{align}
  \widehat{h}^\prime
  = \widehat{h} - {\bm \lambda} \cdot \widehat{\mathbf{t}}
  = \widehat{h} - \lambda_x \widehat{t}_x
  - \lambda_y \widehat{t}_y
  - \lambda_z \widehat{t}_z ,
  \label{AKSE}
\end{align}
where the three Lagrange multipliers $\lambda_x$, $\lambda_y$, and $\lambda_z$ constrain the average values of the three components of the isospin operator. The term ${\bm \lambda} \cdot \widehat{\mathbf{t}}$ 
corresponds to cranking in the isospin space. This procedure is
formally analogous to the cranking model used in high-spin nuclear physics and may be regarded as a first-order approximation to exact isospin projection within the
Kamlah expansion of the projection operator \cite{RingSchuck1980}. In the present case, the cranking term acts in isospin space and approximately projects out the desired
isospin quantum number.

In the spherical polar representation, the isocranking frequencies are expressed as $(\lambda_x,\lambda_y,\lambda_z)=(\lambda \sin\theta \cos\phi,\lambda \sin\theta \sin\phi, \lambda \cos\theta)$.
It has been demonstrated that the results are independent of the azimuthal angle $\phi$ \cite{Sato_2013}, allowing all
calculations to be performed using only the parameters $\lambda$ and $\theta$. This prescription ensures $\langle \widehat{T}_y \rangle = 0$ and preserves the time-reversal symmetry.

In the absence of Coulomb and explicit ISB interactions, neutron-proton mixed DFT preserves exact isospin symmetry and the resulting states possess well-defined
isospin quantum numbers. Consequently, all members of a given isobaric multiplet are degenerate in energy. For intermediate values of $T_z$, the analogue states are 
generated by varying the isocranking angle $\theta$, with $\theta=180^\circ$
corresponding to neutron--proton interchange and $\theta=90^\circ$ corresponds to the $T_z=0$ system, yielding equal neutron--proton mixing. While the degree of mixing varies with $\theta$, the total
energy remains constant in the absence of any ISB terms. The inclusion of the Coulomb interaction lifts this degeneracy and favours states with larger $\langle T_z \rangle$ due to reduced electrostatic energy. 
Thus, at the extremes of the multiplet, 
the lowest-energy states correspond to $T=T_z=|(N-Z)|/2$. In this case, the single-particle energies become $\theta$ dependent and level crossings may occur, potentially leading 
to convergence difficulties. To stabilize the self-consistent iterations, a shifted isocranking prescription is adopted \cite{Sato_2013},
\begin{eqnarray}
{\bm \lambda}
= (\lambda \sin\theta , 0, \lambda \cos\theta)
= (\lambda^\prime \sin\theta^\prime , 0,
\lambda^\prime \cos\theta^\prime + \lambda_{\mathrm{off}}),
\end{eqnarray}
with
\begin{equation}
(\lambda_{\mathrm{off}},\lambda^\prime)
= \frac{1}{2}
\bigl(
\lambda_{np}^{T_z=T}
+
\lambda_{np}^{T_z=-T},
\,
\lambda_{np}^{T_z=T}
-
\lambda_{np}^{T_z=-T}
\bigr).
\label{eq:9}
\end{equation}
Here, $\lambda_{np}$ denotes the neutron-proton Fermi energy difference,
\begin{equation}
  \lambda_{np} = \lambda_n - \lambda_p ,
\end{equation}
where $\lambda_n$ and $\lambda_p$ are the neutron and proton chemical potentials, respectively. In the absence of Coulomb and ISB interactions, $\lambda_{np}$ fixes
the isospin projection $T_z=(N-Z)/2$ of the reference Slater determinant. For the isobaric multiplet with $T_z=T$, the corresponding value
$\lambda_{np}^{(T_z=T)}$ is used as the reference isocranking frequency, while variation of the isocranking angle $\theta$ generates the remaining members of the multiplet.

In addition to the Coulomb interaction, explicit nuclear ISB terms of class II and class III, based on the classification scheme of Henley and Miller of the nucleon-nucleon 
interaction \cite{Henley1969,Miller1990,HenleyMiller1979}, are included in the Skyrme energy density functional to account for
charge-dependent effects beyond the mean-field Coulomb contribution. These terms have been shown to be essential for reproducing mirror displacement energies and
triplet displacement energies in light and medium-mass nuclei \cite{Baczyk2015,Schunck2017,BACZYK2018178}. The corresponding two-body interactions are given by
\begin{eqnarray}
  \widehat{V}^{II}(i,j)
  &=& t_0^{II} \delta(\mathbf{r}_i-\mathbf{r}_j)
  \left(
  3\widehat{\tau}_z(i)\widehat{\tau}_z(j)
  -
  \widehat{\boldsymbol{\tau}}(i)\cdot\widehat{\boldsymbol{\tau}}(j)
  \right), \\
  \widehat{V}^{III}(i,j)
  &=& t_0^{III} \delta(\mathbf{r}_i-\mathbf{r}_j)
  \left(
  \widehat{\tau}_z(i)
  +
  \widehat{\tau}_z(j)
  \right).
\end{eqnarray}
The values of the coupling constants $t_0^{{II}}$ and $t_0^{{III}}$ adopted in the present study are 20 MeV fm$^3$ and $-8$ MeV fm$^3$, respectively \cite{Baczyk2015}. The inclusion of these isospin symmetry breaking terms 
requires a generalized DFT solver incorporating neutron-proton mixing at the Hartree-Fock level.
 
\section{Results and discussion}\label{result}

The isospin-invariant DFT framework which includes an approximate isospin projection technique,
outlined in the preceding section, has been employed to investigate the binding energies and quadrupole properties
for the three selected isobaric triplets of $A=70$, $74$, and $78$. As mentioned before, the motivation for
selecting these three triplets is twofold. First, the $A=70$ triplet represents the earliest reported case of a pronounced
non-linear dependence of the electric quadrupole transition probability, 
$B(E2,2_1^+ \rightarrow 0_1^+)$ on the isospin projection $T_z$ \cite{shape_Wimmer}, challenging the conventional expectation of a linear systematics,
dictated by isospin symmetry. Second, the recent experimental and theoretical investigations 
of the $A=70$, $74$, $78$ triplets have revealed an apparent disparity between the measured quadrupole deformation and the deformation deduced from the
$2^+_1$ excitation energy using the  Grodzins' rule \cite{tension_Henderson}.
\begin{figure}[htbp]
  \centering
  \includegraphics[width=\columnwidth,trim=8 27mm 8 22mm,clip]{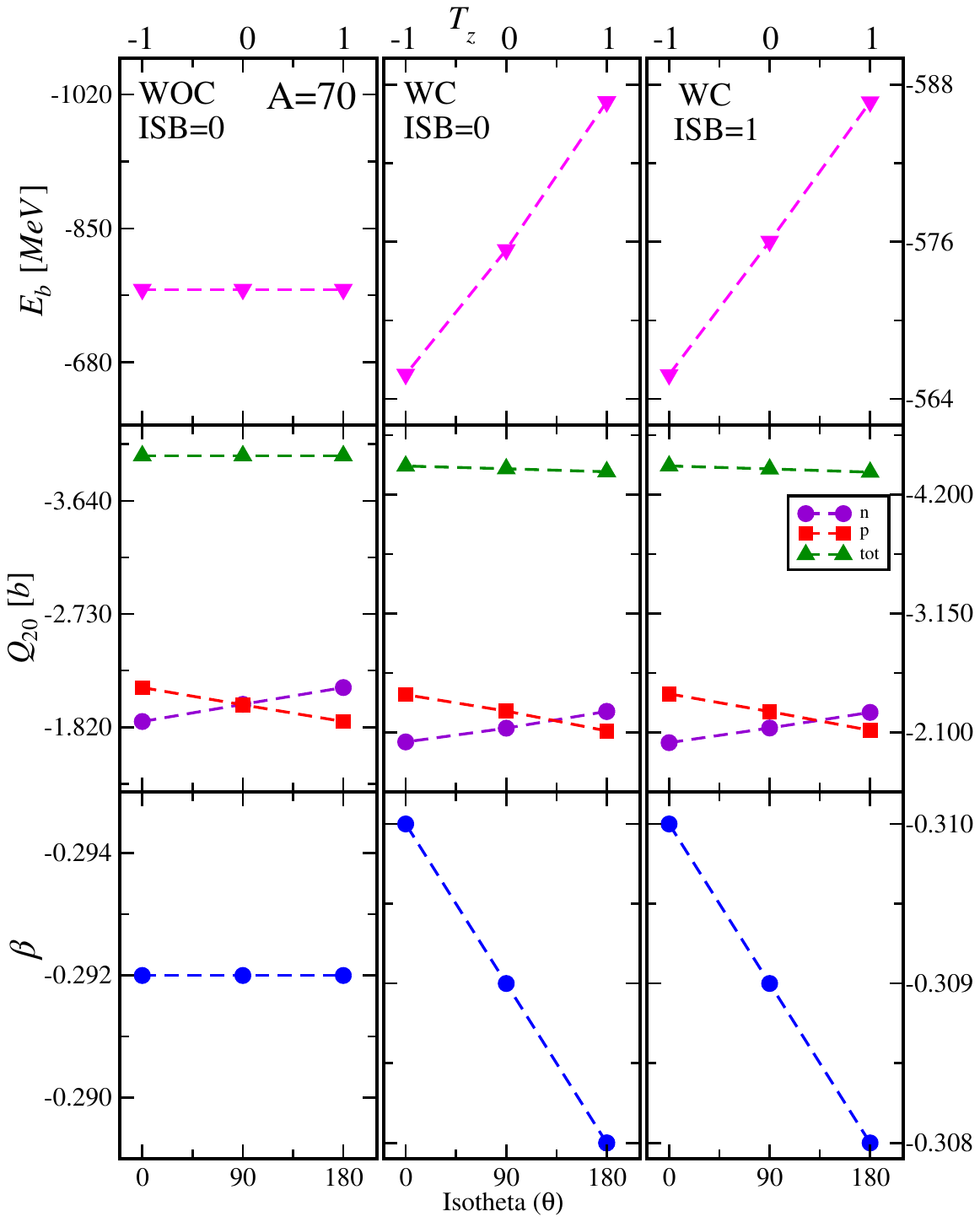}
  \vspace{-18pt}
  \caption{Self-consistently calculated binding energies ($E_b$), quadrupole moments ($Q_{20}$), and quadrupole deformations ($\beta$) for the $A=70$ isospin triplet as 
  functions of rotation angle, or equivalently, of the isospin projection $T_z$.  Three calculation schemes are shown: (i) WOC, ISB = 0, in which both the Coulomb interaction and 
  the additional nuclear isospin symmetry breaking interactions are excluded; (ii) WC, ISB = 0, in which the Coulomb interaction is included but the additional nuclear 
  isospin symmetry breaking interactions are excluded; and (iii) WC, ISB = 1, in which both the Coulomb interaction and the additional nuclear isospin symmetry breaking interactions (class II and class III) are included.} 
  \label{fig:BE2_Tz70}
\end{figure}

In our investigation, we have employed the axial DFT solver HFBTHO \cite{stoitsov2013}, written in cylindrical basis with isospin
mixing \cite{Sheikh_2014} and using the
Skyrme functional $SkM^*$ \cite{Bartel1982}. The main observables extracted from the HFBTHO calculations are the binding energy $``E_b"$ and
the intrinsic quadrupole moments $``Q_{20}^{(q)}"$, defined as,

\begin{equation}
Q_{20}^{(q)} =
\int d^3 r \,
\rho_q(\mathbf{r})
\left(2z^2 - x^2 - y^2\right),
\qquad q=p,n.
\end{equation}

The nuclear shape, characterized by the dimensionless 
quadrupole deformation parameter $\beta$, is obtained through the intrinsic mass quadrupole moment as \cite{BohrMottelson1969}

\begin{equation}
\beta =
\sqrt{\frac{5}{16\pi}}
\frac{4\pi}{3 A R_0^2}
Q_{20},
\qquad
R_0 = r_0 A^{1/3},
\end{equation}
where $r_0 \approx 1.2~\mathrm{fm}$ 
and $A$ is the mass number. 

\begin{figure}[htbp]
  \centering
  \includegraphics[width=\columnwidth,trim=4 27mm 4 22mm,clip]{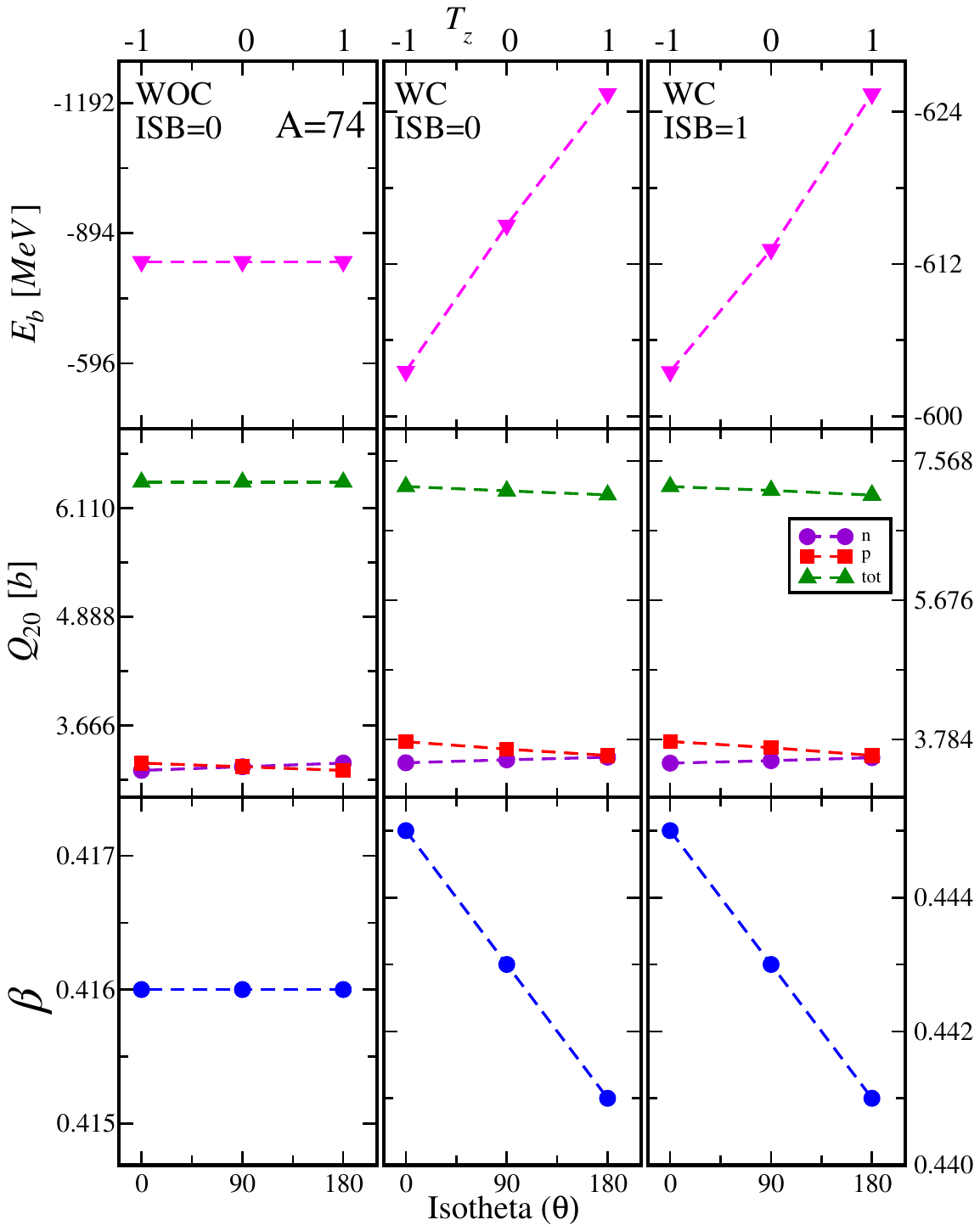}
  \setlength{\abovecaptionskip}{2pt}
  \caption{Self-consistently calculated binding energies ($E_b$), quadrupole moments ($Q_{20}$), and quadrupole deformations ($\beta$) for the $A=74$ isospin triplet as 
  functions of rotation angle, or equivalently, of the isospin projection $T_z$.  Three calculation schemes are shown: (i) WOC, ISB = 0, in which both the Coulomb interaction and 
  the additional nuclear isospin symmetry breaking interactions are excluded; (ii) WC, ISB = 0, in which the Coulomb interaction is included but the additional nuclear 
  isospin symmetry breaking interactions are excluded; and (iii) WC, ISB = 1, in which both the Coulomb interaction and the additional nuclear isospin symmetry breaking interactions (class II and class III) are included.}
  \label{fig:BE2_Tz74}
\end{figure}

\begin{figure}[htbp]
  \centering
  \includegraphics[width=\columnwidth,trim=4 27mm 4 22mm,clip]{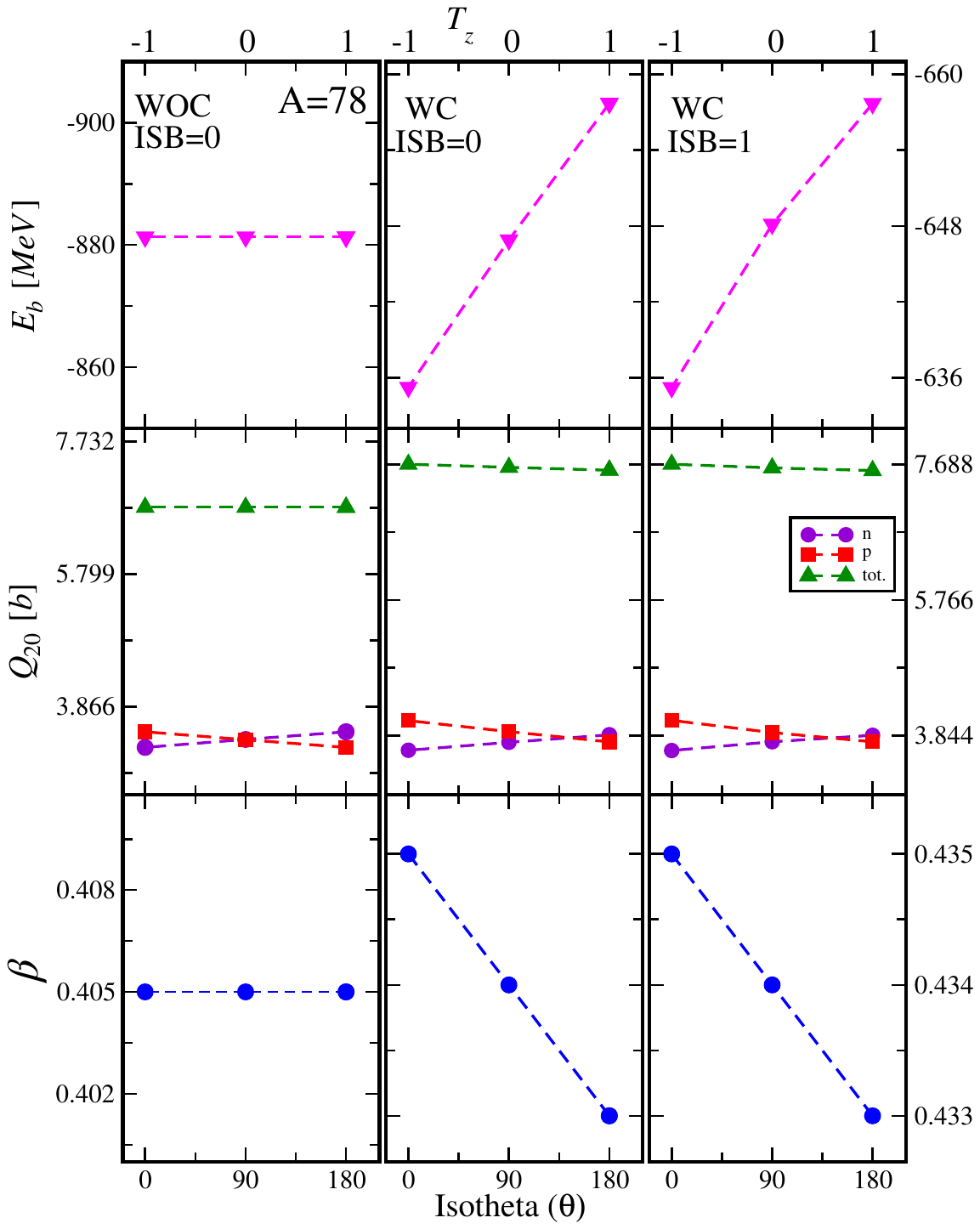}
  \caption{Self-consistently calculated binding energies ($E_b$), quadrupole moments ($Q_{20}$), and quadrupole deformations ($\beta$) for the $A=78$ isospin triplet as 
  functions of rotation angle, or equivalently, of the isospin projection $T_z$.  Three calculation schemes are shown: (i) WOC, ISB = 0, in which both the Coulomb interaction and 
  the additional nuclear isospin symmetry breaking interactions are excluded; (ii) WC, ISB = 0, in which the Coulomb interaction is included but the additional nuclear 
  isospin symmetry breaking interactions are excluded; and (iii) WC, ISB = 1, in which both the Coulomb interaction and the additional nuclear isospin symmetry breaking interactions (class II and class III) are included.
  }
  \label{fig:BE2_Tz78}
\end{figure}

The binding energies and quadrupole deformation properties for the $A=70$, $74$ and $78$ isobaric triplets
are  plotted with respect to the isospin projection $T_z$ or rotation 
angle $\theta$ about the $y$-axis in Figs.~\ref{fig:BE2_Tz70}, \ref{fig:BE2_Tz74}, and \ref{fig:BE2_Tz78}, respectively. Three
sets of
calculations have been 
performed:  ~(i) assuming exact isospin symmetry, (ii) considering isospin symmetry breaking by including the Coulomb interaction 
only and (iii) including the Coulomb interaction together with class~II and class~III isospin-symmetry-breaking terms of the
nuclear force. The corresponding quantities for the three studied isospin triplets are also listed
in Table~\ref{tab:nuclear_combined} in order to depict small changes in these quantities with the inclusion of the isospin breaking terms of the nuclear force.

 \begin{table}[htbp]
\centering
\scriptsize
\setlength{\tabcolsep}{1.8pt}
\caption{Self-consistent calculations of selected nuclear quantities for the isospin triplets with $A = 70$, $74$, and $78$, evaluated at isospin-space orientations of $0^\circ$, $90^\circ$, and $180^\circ$. 
Results are shown both without isospin symmetry breaking terms and with isospin symmetry breaking contributions, including the Coulomb interaction and additional beyond-Coulomb components.
Here $\beta$ denotes quadrupole deformation, $Q_{20}$ denotes quadrupole moment and $E_b$ is the binding energy of a nucleus. Further, WOC, ISB=0 represents the case where the calculation 
was done without the inclusion of Coulomb interaction and Class II and Class III interactions in the nuclear Hamiltonian and so on. }
\label{tab:nuclear_combined}
\resizebox{\columnwidth}{!}{
  \begin{tabular}{l
ccc ccc ccc}
\toprule


Qty
& \multicolumn{3}{c}{WOC ISB = 0}
& \multicolumn{3}{c}{WC ISB = 0}
& \multicolumn{3}{c}{WC ISB = 1} \\

& $0^\circ$ & $90^\circ$ & $180^\circ$
& $0^\circ$ & $90^\circ$ & $180^\circ$
& $0^\circ$ & $90^\circ$ & $180^\circ$ \\

\midrule
\multicolumn{10}{c}{\textbf{A = 70}} \\
\midrule

$\beta$
& -0.292 & -0.292 & -0.292

& -0.310 & -0.310 & -0.308
& -0.310 & -0.309 & -0.308 \\

$Q_{20}^{(n)}$
& -1.865 & -2.004 & -2.138

& -2.017 & -2.138 & -2.286
& -2.011 & -2.140 & -2.278 \\

$Q_{20}^{(p)}$
& -2.138 & -1.999 & -1.865

& -2.434 & -2.289 & -2.114
& -2.441 & -2.285 & -2.119 \\

$Q_{20}^{(tot)}$
& -4.003 & -4.003 & -4.003

& -4.451 & -4.427 & -4.400
& -4.452 & -4.425 & -4.397 \\

$E_{b}$
& -772.473 & -772.473 & -772.473

& -565.872 & -575.389 & -586.690
& -565.853 & -576.032 & -586.673 \\

\midrule
\multicolumn{10}{c}{\textbf{A = 74}} \\
\midrule

$\beta$
& 0.416 & 0.416 & 0.416
& 0.445 & 0.443 & 0.441
& 0.445 & 0.443 & 0.441 \\

$Q_{20}^{(n)}$
& 3.160 & 3.202 & 3.243

& 3.467 & 3.507 & 3.542
& 3.462 & 3.495 & 3.536 \\

$Q_{20}^{(p)}$
& 3.243 & 3.202 & 3.160

& 3.752 & 3.653 & 3.565
& 3.756 & 3.672 & 3.566 \\

$Q_{20}^{(tot)}$
& 6.403 & 6.403 & 6.403

& 7.219 & 7.160 & 7.106
& 7.219 & 7.167 & 7.102 \\

$E_{b}$
& -828.991 & -828.991 & -828.991

& -603.543 & -615.058 & -625.415
& -603.521 & -613.138 & -625.396 \\

\midrule
\multicolumn{10}{c}{\textbf{A = 78}} \\
\midrule

$\beta$
& 0.405 & 0.405 & 0.405

& 0.435 & 0.434 & 0.433
& 0.435 & 0.434 & 0.433 \\

$Q_{20}^{(n)}$
& 3.272 & 3.383 & 3.500

& 3.635 & 3.747 & 3.853
& 3.631 & 3.754 & 3.847 \\

$Q_{20}^{(p)}$
& 3.500 & 3.384 & 3.272

& 4.055 & 3.899 & 3.753
& 4.059 & 3.885 & 3.755 \\

$Q_{20}^{(tot)}$
& 6.772 & 6.772 & 6.772

& 7.690 & 7.647 & 7.606
& 7.690 & 7.639 & 7.601 \\

$E_{b}$
& -881.23 & -881.23 & -881.23

& -635.238 & -646.893 & -657.702
& -635.216 & -648.131 & -657.682 \\

\bottomrule
\end{tabular}
}

\end{table}

In the limit of exact isospin symmetry, the binding energies and the total quadrupole deformations of the
 studied isospin triplets are 
identical (see Figs. \ref{fig:BE2_Tz70},  \ref{fig:BE2_Tz74} and  \ref{fig:BE2_Tz78}) for all three members in each case, as required by their invariance under rotations in isospin space.
When the Coulomb interaction is included in the density functional, the binding energy increases as we go from $T_z=-1 (\theta=0^\circ)$ to
$T_z=+1 (\theta=180^\circ)$. This is expected as the protons are replaced by neutrons and the Coulomb energy is diminished. 
The deformation parameter, $\beta$ is noted to drop slightly with increasing $T_z$.
The small changes in the calculated quantities resulting from the inclusion of the additional terms of class~II and class~III
are not discernible in Figs. \ref{fig:BE2_Tz70},  \ref{fig:BE2_Tz74} and  \ref{fig:BE2_Tz78}. 
However, the subtle variations of these quantities with $T_z$ are noticeable in Table~\ref{tab:nuclear_combined}.

In the rigid-rotor approximation of the collective model, the rotational excitation energy is inversely proportional to the moment of inertia which, in turn,
depends directly on the quadrupole deformation of a nucleus \cite{BohrMottelson1969}.
Since the deformations are almost equal across each isospin triplet, the corresponding $E_{2_1^+}$ energies shall also be nearly identical in each case.
This is approximately the case for the three isospin triplets as can be seen from Fig. 1
of the Ref. \cite{tension_Henderson}. Most importantly, it is evident from our calculations that the three members in each of the studied isospin triplets exhibit identical quadrupole shapes.

The intrinsic proton quadrupole moments, $Q_{20}^{(p)}$, obtained from the DFT calculations for the members of the three 
isospin triplets were used to evaluate
$\alpha |Q_{20}^{(p)}|^2$. The constant $\alpha$ was fixed by requiring this quantity to reproduce
the experimental
$B(E2;2_1^+ \rightarrow 0_1^+)$ value for the $T_z=+1$ member in each case, including all isospin-breaking contributions from the Coulomb interaction
and the ISB terms.

\begin{figure}[htbp]
  \centering
  \includegraphics[width=\columnwidth,trim=4 25mm 4 20mm,clip]{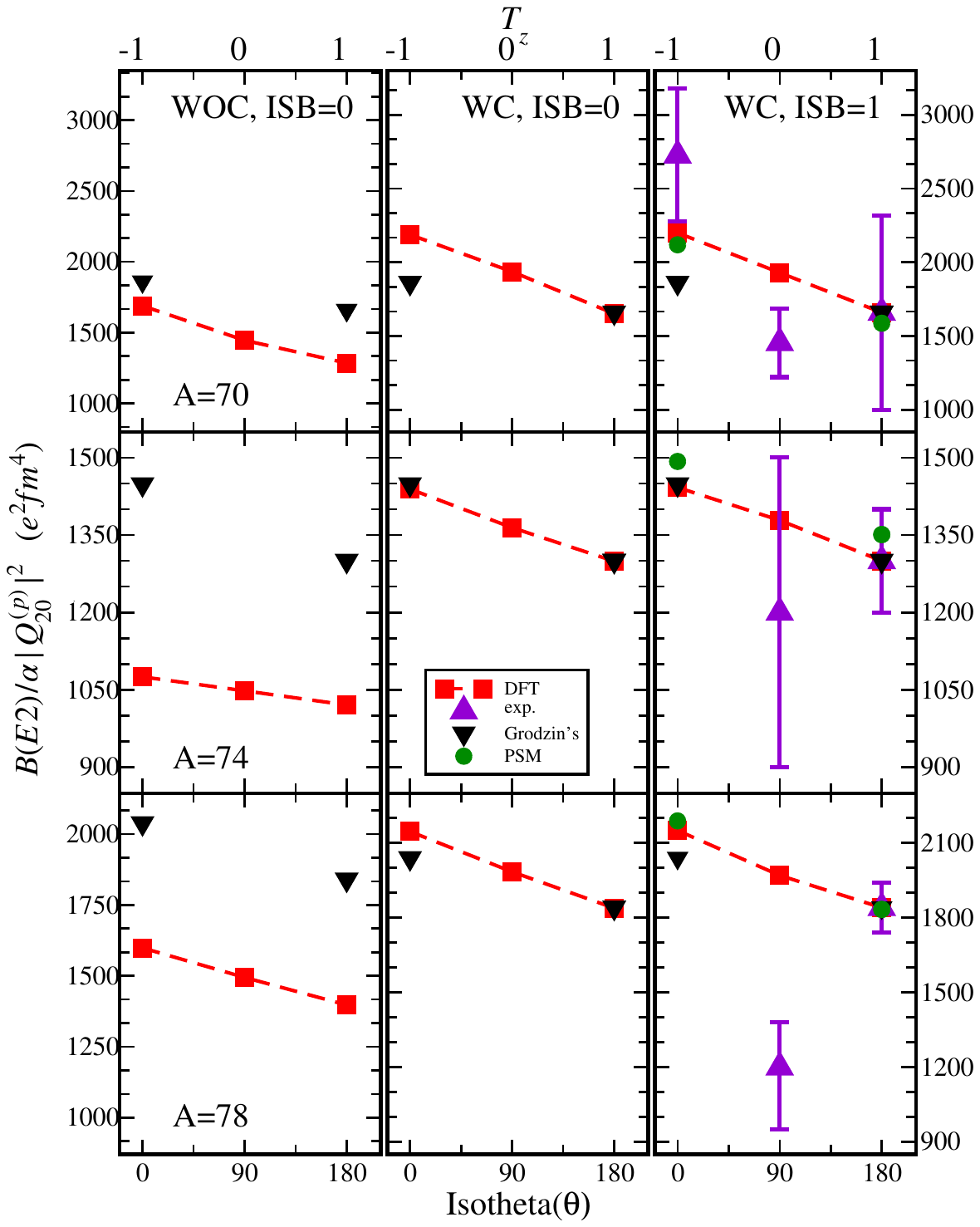}
  \caption{Behavior of the $B(E2,2_1^+ \rightarrow 0_1^+)$ rates obtained from the self-consistently calculated quadrupole charge moments for the isospin triplets with $A=70, ~74$ and $78$,  as a function of rotation 
  about the $y$-axis in isospin space, or equivalently,  the isospin projection $T_z$. The calculated values of $B(E2,2_1^+ \rightarrow 0_1^+)$ using PSM calculations and the Grodzins' rule for the $T_z=\pm 1$ are also shown. 
  The calculated values from DFT and Grodzins' rule were normalized with the experimental values of the $T_z=+1$ members for the comparison of the experimental and the theoretical trends. The experimental values were taken from \cite{Pritychenko2014,Llewellyn2020,Morse2018,shape_Wimmer}.}
  \label{fig:70BE2_Tz}
\end{figure}

The above comparison is based on the fact that the $B(E2;2_1^+ \rightarrow 0_1^+)$ strength is determined by the charge quadrupole
transition matrix element, which is primarily governed by the proton distribution \cite{Dobaczewski2007}. Thus,
$\alpha |Q_{20}^{(p)}|^2$ provides a meaningful measure of the calculated quadrupole strengths that can be compared with the experimental
$B(E2;2_1^+ \rightarrow 0_1^+)$ systematics within each isospin triplet. These results for the three triplets are shown in Fig.~\ref{fig:70BE2_Tz}. In the limit
of exact isospin symmetry, the $B(E2;2_1^+ \rightarrow 0_1^+)$ strengths in each triplet are expected to decrease  linearly with increasing $T_z$, as shown
in the left panel of the figure.  The inclusion of isospin-breaking interactions produces only minor changes in this trend, as shown
in the middle and right panels.

For the three studied isospin triplets, the experimental $B(E2,2_1^+ \rightarrow 0_1^+)$ rates are available for all the three members of $A=70$ triplet only. We emphasize here that the 
calculated $B(E2,2_1^+ \rightarrow 0_1^+)$ rate for the $T_z=0$ member of this triplet, namely ${}^{70}_{35}\mathrm{Br}$, overestimates the observed value leading to the non-linear 
trend of the measured $B(E2,2_1^+ \rightarrow 0_1^+)$ rates of the $A=70$ triplets. The decreasing trend of $B(E2,2_1^+ \rightarrow 0_1^+)$ with $T_z$ is consistent with the 
experimentally measured values of the $T_z=\pm1$ members of this triplet. 
Thus, the focus of any future studies trying to understand the non-linearity of $B(E2,2_1^+ \rightarrow 0_1^+)$ rates in the $A=70$ triplet is
the anomalous behaviour of the $T_z=0$ member. It is noted from Fig.~\ref{fig:70BE2_Tz} that the DFT calculated values
overestimate the experimentally observed $B(E2,2_1^+ \rightarrow 0_1^+)$ 
values for the $T_z=0$ members of the $A=$74 and 78 isospin triplets as well.

We also plotted the estimated $B(E2,2_1^+ \rightarrow 0_1^+)$ rates calculated using the Grodzins' rule for
comparison between the $T_z=\pm 1$ members in each triplet. This rule essentially uses the transition energy of the
$2^+_1$ state for evaluating the $B(E2,2_1^+ \rightarrow 0_1^+)$ rates in even-even 
systems. These rates were also normalized with the experimental values of the $T_z=+1$ member in each studied case.
They also agree with the trend of our results predicting an enhanced $B(E2,2_1^+ \rightarrow 0_1^+)$ rate for the $T_z=-1$ member as compared to the
$T_z=+1$ member. 

We would like to point out that the above calculated $B(E2,2_1^+ \rightarrow 0_1^+)$ rates from the Skyrme DFT approach are approximate. In an accurate treatment, the transition rate needs to be
calculated from states with well defined angular momentum. This would require angular momentum projected HFBTHO approach, which
has singularity problems as discussed in ref.~\cite{Sheikh2021}. In order to evaluate the corrections to the calculated $B(E2,2_1^+ \rightarrow 0_1^+)$ rates from the angular momentum projected states,
we have performed projected shell model (PSM) calculations \cite{Hara1995,Sun1994,Palit2001,xbsc-q21d} . For these calculations, the quadrupole deformation parameters needed in the PSM approach were used
from the self-consistent Skyrme-DFT calculations. The PSM calculations were performed only for $T_z=\pm 1$ as for the odd-odd
system having $T_z=0$, neutron-proton pairing needs to be included to obtain the $I=0$ paired ground state. It is known that for the $T_z=0$
member of each triplet, the
ground-band is an $I=0$  $(T=1)$ paired configuration as for the even-even systems \cite{Jenkins2002Br70,Fischer2006Rb74,Zimba2025Zr78}.

As is evident from Fig.~\ref{fig:70BE2_Tz}, PSM results exhibit the same qualitative
isospin dependence of the $B(E2,2_1^+ \rightarrow 0_1^+)$ rates as a function of $T_z$ as those obtained from  the isospin-invariant Skyrme-DFT calculations and the Grodzins' rule. 
Hence, it is important to note that none of the calculations performed on these nuclei
imply any discrepancy between the $B(E2,2_1^+ \rightarrow 0_1^+)$ rates predicted by isospin symmetry and the quadrupole deformation as
highlighted in Ref. \cite{tension_Henderson}. The reported discrepancy arises by extrapolating the experimentally measured values of
$T_z=0$ and $+1$ members. A plausible explanation for the suppressed $B(E2,2_1^+ \rightarrow 0_1^+)$ rates in the $T_z=0$ members of the 
three studied isospin triplets is the mixing between the states $|T=1,T_z=0\rangle$ and $|T=0,T_z=0\rangle$. However, this possibility could not 
be investigated in the present work, as it would require three-dimensional isospin projection. It is quite evident from the present study that the
approximate isospin projection performed is inadequate for the $T_z=0$ members of the triplets.

The shape change between the $T_z=\pm 1$ members was proposed in Ref. \cite{shape_Wimmer} as a possible explanation for the non-linearity 
observed in the $A=70$ triplet. However, this possibility has been ruled out in a recent study \cite{Backes2025}. Furthermore, our calculations suggest 
a common oblate shape for the $A=70$ isospin triplet. This is evident from the quadrupole deformation $(\beta)$ values listed in Table~\ref{tab:nuclear_combined} and plotted in Fig. \ref{fig:BE2_Tz70}.
\begin{table}[!t]
\centering
\caption{The Lagrange  multipliers used in the present calculations. These multipliers were calculated using the Eq. (7) above. 
The calculations were done with $10$  major oscillator shells.}
\label{tab:lagrange}
\setlength{\tabcolsep}{5pt}
\renewcommand{\arraystretch}{1.05}

\resizebox{\columnwidth}{!}{
\begin{tabular}{c cc cc cc}
\toprule
& \multicolumn{2}{c}{\textbf{WOC, ISB = 0}}
& \multicolumn{2}{c}{\textbf{WC, ISB = 0}}
& \multicolumn{2}{c}{\textbf{WC, ISB = 1}} \\

\cmidrule(lr){2-3}
\cmidrule(lr){4-5}
\cmidrule(lr){6-7}

\textbf{Case}
& \textbf{$\lambda'$}
& \textbf{$\lambda_{\mathrm{off}}$}
& \textbf{$\lambda'$}
& \textbf{$\lambda_{\mathrm{off}}$}
& \textbf{$\lambda'$}
& \textbf{$\lambda_{\mathrm{off}}$} \\
\midrule

\textbf{70} & -2.7826 & 0.4774 & 3.2049 & -10.7531 & 3.1430 & -11.0236 \\
\textbf{74} & -2.6497 & 0.0046 & 3.1395 & -10.7954 & 2.6435 & -11.7985 \\
\textbf{78} & 2.8876 & 0.0553 & 2.5067 & -11.1590 & 2.6464 & -11.3013 \\

\bottomrule
\end{tabular}
}
\end{table}

It is worth noting that the error bar in the available measurement of $B(E2,2_1^+ \rightarrow 0_1^+)$ for the $T_z=0$ member
of the $A=74$ triplet is so large that the calculated value falls within it. Thus, the linear 
variation of $B(E2,2_1^+ \rightarrow 0_1^+)$ rates in this case cannot be completely ruled out. Similarly, our calculations
predict a prolate shape for the triplets with almost identical quadrupole deformation, both for $A=74$ and $A=78$ (Figs.  \ref{fig:BE2_Tz74} and  \ref{fig:BE2_Tz78}). However, it is important to notice
that the measured $B(E2,2_1^+ \rightarrow 0_1^+)$ rate for the $T_z=0$ member of the $A=78$ triplet is significantly suppressed as compared to the calculated values.


It is clear from the comparisons shown in Table~\ref{tab:nuclear_combined}
that the ISB terms introduce minor corrections, especially for nuclei 
with $T_z=\pm 1$. These corrections are negligible in comparison to the isospin mixing due to the Coulomb interaction. 
The numerical values of the Lagrange multipliers used in the self-consistent minimization procedure for the three isospin triplets are listed in
Table~\ref{tab:lagrange} and it is evident that ISB terms lead to small corrections in these quantities.

 \section{Summary and conclusions}\label{sum}

In the present work, we have performed a systematic microscopic investigation of the interplay between the isospin symmetry and the quadrupole collectivity
 in the mass region $A\approx70$-80 using the isospin-invariant 
 nuclear Skyrme density functional approach. Fully self-consistent Hartree-Fock calculations, followed by the
 approximate isospin projection, have been carried out for the  $T=1$ isobaric triplets with $A=70$, 74, and 
 78, incorporating the Coulomb interaction and class~II and class~III isospin-symmetry-breaking terms of the nuclear force.
 We also performed the PSM calculations for the even-even systems of the studied nuclei.

 The principal objective of the present study was to understand the anomalous systematics observed in the reduced electric quadrupole transition
 probabilities, $B(E2; 2_1^+ \rightarrow 0_1^+)$, in some nuclei in the $A\approx70$-80 region. This region has been the subject of intense
 studies recently in order to understand the observed non-linear trend of the $B(E2,2_1^+ \rightarrow 0_1^+)$ rates in $A=70$ triplets and the behaviour of
 these rates in other triplets, particularly those corresponding to $A=74$ and $78$. Shape change was argued to be the reason of the observed
 non-linear trend in the $A=70$ triplets \cite{shape_Wimmer}. However, a detailed study of even-even mirror nuclei in triplets with $42\leq A \leq 98$  range
 was performed, and the shape change in mirror nuclei of the above three triplets was ruled out \cite{Backes2025}.

 Our calculations based on isospin-invariant density functional theory also indicate  uniform shapes across each of the three studied isospin triplets. In particular, the
 triplets corresponding to $A=70$ are  expected to be oblate with almost equal quadrupole deformations. Similar conclusions were also derived for the
 triplets corresponding to $A=74$ and $78$ with 
 prolate shapes. The essential observation of our study is the suppressed measured $B(E2,2_1^+ \rightarrow 0_1^+)$
 rates for the $T_z=0$ members as compared to the calculated values, and estimates based on isospin symmetry. These consistently suppressed
 $B(E2,2_1^+ \rightarrow 0_1^+)$ rates in the $T_z=0$ members of the studied isospin triplets are the actual reason for the non-linear trend expected in all of
 the three triplets. To further understand the anomalous behaviour of the $B(E2,2_1^+ \rightarrow 0_1^+)$ rates in the three studied isospin triplets, the lifetime 
 measurements of the $T_z=-1$ members of the $A=74$ and $78$  triplets are very crucial. On the theoretical side, three-dimensional isospin projection ought to be performed within the isospin
 invariant formalism that will project out the $T$=0 and 1 states of the $T_z=0$ nuclei. Further neutron-proton pairing is expected to be
 important for $N=Z$ systems, and needs to be included in the DFT approach to investigate its effects on the quadrupole
 collectivity.

\section*{Acknowledgements}

S.T. acknowledges the Department of Science and Technology (Government of India) for the award of INSPIRE fellowship under Sanction No. DST/INSPIRE Fellowship/[2023/IF230401].

\bibliographystyle{apsrev4-2}
\bibliography{references_ordered}
\end{document}